\documentclass[12pt]{article}
\usepackage{geometry}
\usepackage{array}
\usepackage{longtable}
\usepackage{afterpage}
\usepackage{pdflscape}
\usepackage{arydshln}
\usepackage{multirow}
\usepackage{float}
\usepackage{url}
\usepackage{amsmath}
\usepackage{amsthm}
\usepackage{eufrak}
\usepackage{bbm}
\usepackage{graphicx}
\usepackage{esint}
\usepackage{setspace}
\usepackage{titlesec}
\usepackage{color}
\usepackage[acronym,nonumberlist]{glossaries}
\definecolor{Blue}{rgb}{0.0, 0.0, 1}
\definecolor{Green}{rgb}{0.0, 1, 0.0}
\usepackage{natbib}
\titleformat{\section}{\large\bfseries}{\thesection}{1em}{}
\titleformat{\subsection}{\bfseries}{\thesubsection}{1em}{}
\titleformat{\subsubsection}{\itshape}{\thesubsubsection}{1em}{}

\newcommand{\Keywords}[1]{\par\noindent
{\small{\em Keywords\/}: #1}}

\begin{document}

\title{\Large Tail Risk Premia for Long-Term Equity Investors}

\author{\large Johannes Rauch and Carol Alexander\footnote{School of Business, Management and Economics, University of Sussex, UK. c.alexander@sussex.ac.uk;  j.rauch@sussex.ac.uk.}}

\date{This version: January 2016}
\maketitle

\doublespacing
\thispagestyle{empty}
\begin{abstract}\normalsize
\noindent  
We use the P\&L on a particular class of swaps, representing variance and higher moments for log returns, as estimators in our empirical study on the S\&P500 that investigates the factors determining variance and higher-moment risk premia. This class is the discretisation invariant sub-class of swaps with Neuberger's aggregating characteristics. Besides the market excess return, momentum is the dominant driver for both skewness and kurtosis risk premia, which exhibit a highly significant negative correlation.  By contrast, the variance risk premium responds positively to size and negatively to growth, and the correlation between variance and tail risk premia is relatively low compared with previous research, particularly at high sampling frequencies. These findings extend prior research on determinants of these risk premia. Furthermore, our meticulous data-construction methodology avoids unwanted artefacts which distort results.  \\
\end{abstract}

\newpage

\newtheorem{thm}{Theorem}[section]
\newtheorem{pro}[thm]{Proposition}
\newtheorem{con}[thm]{Conjecture}
\newtheorem{lem}[thm]{Lemma}
\theoremstyle{definition}
\newtheorem{den}[thm]{Definition}
\newtheorem{prf}[thm]{Proof}
\newtheorem{rem}[thm]{Remark}

\newacronym{ap}{AP}{Aggregation Property}
\newacronym{erp}{ERP}{equity risk premium}
\newacronym{iid}{i.i.d.}{independent and identically distributed}
\newacronym{lv}{LV}{log variance}
\newacronym{otm}{OTM}{out-of-the-money}
\newacronym{pnl}{P\&L}{profit and loss}
\newacronym{qv}{QV}{quadratic variation}
\newacronym{rv}{RV}{realised variance}
\newacronym{spx}{SPX}{Standard \& Poor's 500 stock market index}
\newacronym{trp}{TRP}{tail risk premium}
\newacronym{vix}{VIX}{CBOE Volatility Index}
\newacronym{vrp}{VRP}{variance risk premium}

\setlength\abovedisplayskip{6pt}
\setlength\belowdisplayskip{6pt}
\setlength\abovedisplayshortskip{0pt}
\setlength\belowdisplayshortskip{6pt}

\clearpage
\setcounter{page}{1}

\noindent A risk premium is the amount that an investor requires to compensate for taking a risk. Pioneered by \cite{CW09}, recent research has focused on modelling, measuring and analysing the determinants of the \gls{vrp}. 
An investor can hold variance by taking a delta-hedged options position, but since the 1990's a purer form of variance has been accessible through over-the-counter trades on variance swaps. Either way an estimator for the \gls{vrp} is the difference between the expected value of some second moment characteristic under the physical measure and its expectation under the risk-neutral measure.

Two main strands of literature have emerged regarding the estimation of the \gls{vrp}. The first, exemplified by the work of \cite{BCJ07} and many others since, estimate a stochastic volatility-jump model using both option prices and the underlying time series. The advantage of this approach is that it provides a direct estimate of the risk premium as a function of the model parameters estimated under the physical and risk-neutral measures. A disadvantage of this `model-dependent' approach is that results depend on the specification of the model for the underlying price and volatility processes, and the suitability of most specifications depends on the market conditions over the sample period.\footnote{To facilitate the risk-neutral model calibration this is typically assumed to be affine. But non-affine volatility processes (e.g. GARCH) which allow volatility to be more flexible than it is in a square-root process reduces the need for jumps and are arguably  better specified for equity index dynamics. See \cite{KA13} and others.} Most papers, with the exception of \cite{BGL13} and \cite{AKM14}, focus only on short-term risk premia. Notably, \cite{T10} and \cite{BT11} model the tail risk that is driven by jumps in the price and volatility processes. Using high-frequency data and short maturity out-of-the money option prices they demonstrate that jumps account for a large fraction of the short-term \gls{vrp} in the S\&P500.

Understanding the factors driving the variance and tail risk premia is important for asset pricing, trading, and investment management; but risk premia are notoriously difficult to estimate. While short-term tail risk is mainly driven by jumps, it is serial dependence that dominates the long-term. Our work develops a methodology to identify the factors which drive long-term tail risk. As such it falls squarely within the other, `model-free' branch of the risk premium literature. This takes its roots from the identification by \cite{BK03} of the \gls{vrp} with delta-hedged gains on options, the model-free variation measures introduced by \cite{CW03} and the subsequent work of \cite{BM06} linking the \gls{vrp} with higher moments of returns. Following \cite{CW09} this strand of the literature models the \gls{vrp} as the expected \gls{pnl} on a swap which exchanges a realised variance for a model-free swap rate.\footnote{In fact, increasingly the exposure to a \gls{vrp} is becoming available via transactions on exchange-traded products. For example, instead of buying a variance swap on the S\&P500 index, the investor could purchase a \gls{vix} futures contract. The two exposures are not perfect substitutes but they are highly correlated at very short-term horizons, making such products interesting for speculators. At longer horizons, investments are heavily eroded by roll costs on a term structure that is almost always in contango. See \cite{AKK15} for a summary of trading and statistical characteristics and a recent literature review of volatility products.}
The advantages of the model-free approach are that results do not depend on any particular process dynamics for the underlying price and volatility, that \gls{pnl} time series for a swap (i.e. estimators for the risk premium) can be generated for various investment horizons, sampling intervals and monitoring frequencies, and that these series may be used to examine the factors which determine risk premia under changing market conditions. 

A fundamental problem in asset pricing is to capture codependencies between variance and tail risk premia by understanding their common drivers. Our empirical study will focus on the S\&P500 at different investment horizons between 1 and 6 months. For very short-term risk premia the methodology of choice may be to forecast the entire return distribution, for horizons of 1 month or more almost all the literature focuses only on its moments, usually the variance but also the skewness.\footnote{In the S\&P500 market `short-term' refers to horizons of days, even hours. The S\&P500 derivatives markets are among the most active in the world, with the average holding time of a standard futures being a matter of minutes, and volatility contracts rarely being held for more than one week.} However, long-term (moment) risk premia are notoriously difficult to estimate for two main reasons: firstly, statistical estimators of risk premia may be biased and/or inefficient -- arguably the worst case would be for an estimator to be biased and efficient because the estimates are more likely to be wrong -- and secondly, at long-term horizons there are insufficient time-series data available for reliable estimation.\footnote{To obtain a sufficient degree of statistical accuracy (with non-overlapping sampling) one would need to look back to the 1980's or before, when financial markets had very different characteristics than they do today. Indeed, so much has changed during the last decade that even using monthly data over a 10-year period -- giving a statistically minimal 120 observations -- would require robustness checks (e.g. during the 2008--9 banking crisis) that we simply do not have the data to perform.}

To our knowledge, our paper is the first to provide a sound theoretical framework for conducting a large-sample tail risk premium analysis. We investigate how the bias and efficiency of the statistical estimator for a long-term risk premium depend on the martingale assumption, on the investment horizon and sampling frequency for the estimate, and on the choice of the realised pay-off. However, we regard the following as our
 main contributions so far: First, we analyse kurtosis as well as variance and skewness risk premia; Second, we develop a novel methodology for constructing non-overlapping, investable excess returns on moment swaps by  decomposing these into an unbiased implied component which determines the risk premium and a realised component which controls the precision
of the risk premium estimator; Thirdly, by estimating variance, skewness and kurtosis risk premia at different sampling frequencies we present new empirical evidence regarding their determinants. No other papers (to our knowledge) examine the common drivers of variance, skewness and kurtosis risk premia, and the conclusion that S\&P500 variance and skewness risk premia are very highly and negatively correlated (which would indicate that it is not possible to make systematic profit from a delta-vega-hedged options position) is based on questionable methodological 
assumptions. In contrast with previous empirical work, notably \cite{CW09}, we find a strong asymmetric response of all risk premia to market shocks especially when measured at high sampling frequency. Further, we can distinguish the driving factors of the \gls{vrp} from the factors determining the tail risk premia, concluding that long-term investors can indeed diversify their exposures to variance risk (which is driven largely by growth) and skewness or kurtosis risk (which are highly correlated  and are driven largely by momentum).

Besides furthering much previous academic literature, our work has considerable relevance for practitioners. Understanding the common determinants of skewness and kurtosis risk premia, and re-examining their relationship with the \gls{vrp}, is particularly relevant for risk-averse investors and for the banks which provide variance and skewness swaps.  An additional advantage from the practitioner's perspective is that our class of pay-offs allows for the construction of instruments that can be exactly replicated by the issuer (and hence priced very tightly) and which enable the user to precisely hedge or gain exposure to tail risk.

In the following: Section \ref{sec:background} explains the challenge of inferring long-term from short-term higher moments and then sets our research questions within the framework of the  literature on measuring and examining the factors determining the VRP in the model-free (i.e. swaps-based) setting; 
Section \ref{sec:theory} provides notation with a brief overview of variance swaps, and of the \gls{ap} introduced by \cite{N12}, and the more restricted class of Discretisation-Invariant (DI) contracts developed in \cite{AR16} which satisfy the AP;
A new theory is presented in Section \ref{sec:LTTRP} which explains how to construct the investable \glspl{pnl} on moment swaps when  estimating tail risk premia at different sampling frequencies. The section concludes with a characterisation of bias and efficiency in moment risk premium estimators; 
The empirical study on the S\&P500 index, using 18 years of daily closing price data is presented in Section \ref{sec:empirical}. Here we construct ex-ante estimates for moment risk premia sampled at daily, weekly and monthly frequencies. Then we analyse the common and idiosyncratic Fama-French and Carhart factors (see \cite{FF93} and \cite{C97}) which determine them, for a 30-day investment horizon;\footnote{Results for investment horizons of between 1 and 6 months are available on request. They are omitted for brevity and because the new insights on term-structure risk premia analysis that the results provide are not of major relevance to this paper.} Section \ref{sec:conc} concludes. All proofs are in the Appendix. 

\section{Background}\label{sec:background}
\subsection{Inferring Long-term from Short-Term Risk Premia}
When is it possible to infer the values for long-term moments from the information contained in short-term returns? Table \ref{tab:returns} illustrates the issue by displaying the usual sample standard deviation, skewness and excess kurtosis of the daily, weekly and monthly changes in S\&P500 futures prices over three separate 6-year periods. In each case the weekly and monthly observed moment is compared with the moment inferred under the assumption of \gls{iid} daily returns, where the standard deviation scales with the square root of the horizon, skewness scales with the inverse square root and excess kurtosis scales with the inverse of the horizon. 

\begin{table}[h!]\small
\vspace{10pt}
\begin{center}
\begin{tabular}{ll|c:cc:cc}
\multicolumn{2}{l}{Return (trading days)} & daily (1) & \multicolumn{2}{c}{weekly (5)} & \multicolumn{2}{c}{monthly (20)} \\
& & obs. & obs. & \gls{iid} & obs. & \gls{iid}\\ \hline
\multirow{3}{*}{Standard deviation} & '96--'01 & 0.013 & 0.025 & 0.028 & 0.051 & 0.056 \\
& '02--'07 & 0.010 & 0.021 & 0.022 & 0.034 & 0.045 \\
& '08--'13 & 0.015 & 0.032 & 0.034 & 0.064 & 0.069 \\ \hdashline
\multirow{3}{*}{Skewness} & '96--'01 & -0.16 & -0.02 & -0.07 & 0.18 & -0.04 \\
& '02--'07 & 0.12 & -0.02 & 0.05 & -0.41 & 0.03 \\
& '08--'13 & 0.34 & -0.63 & 0.15 & -1.20 & 0.08 \\ \hdashline
\multirow{3}{*}{Excess kurtosis} & '96--'01 & 2.79 & 1.69 & 0.56 & 0.72 & 0.14 \\
& '02--'07 & 3.01 & 2.16 & 0.60 & 0.55 & 0.15 \\
& '08--'13 & 12.1 & 10.0 & 2.42 & 6.17 & 0.60
\end{tabular}
\end{center}
\vspace{-10pt}
{\caption[Aggregation]{\small Sample statistics of S\&P500 futures: observed vs. predicted under \gls{iid}}\label{tab:returns}}
\end{table}

\noindent The observed values for weekly and monthly standard deviation are roughly the same as those inferred using the \gls{iid} assumption; however, the same cannot be said about skewness and excess kurtosis. During the more stable 1996--2001 period observed skewness is negative for daily returns but it becomes positive for monthly returns. However, the \gls{iid} assumption predicts that skewness remains negative while decreasing in absolute terms. The opposite pattern holds during the more turbulent periods from 2002--2007 and 2008--2013: here the observed skewness is positive for daily returns and highly negative for weekly and monthly returns. This confirms the introductory statement of \cite{N12} that, far from diminishing with horizon as would be the case if returns were \gls{iid}, skewness actually increases (in absolute terms) with horizon. Our novel finding here is the discrepancy between the \gls{iid} assumption and observed returns when conducting the same analysis for excess kurtosis: over both sample periods, observed excess kurtosis decreases much slower with horizon than \gls{iid} would predict. 

\subsection{Drivers of Long-term vs. Short-Term Risk Premia}

The main driver of short-term tail risk is jumps in the underlying; these influence kurtosis and, if asymmetric, also skewness. For the long-term horizon, skewness is mainly driven by the negative correlation between price and variance, and kurtosis results from volatility clustering. Therefore, accurate information about serial dependence is essential in order to infer long-term moments from short-term data. However, among the many choices that must be made when constructing a time series for the realised \gls{vrp}, some authors have introduced autocorrelation as an unwanted artefact of the data construction methodology. Decision variables include the investment horizon, the underlying asset, the sample period, the realised pay-off, the monitoring partition and the sampling frequency. The first three variables are simply a matter for individual preferences regarding the empirical insights of relevance. We shall focus on choices made about the latter three variables because they have very important modelling implications for risk premium estimation:

\begin{enumerate}

\item Realised pay-off: For a variance swap this is typically the sum of the squared log returns. However, using such a pay-off introduces errors to the model-free swap rate formula, so that the market rates can deviate substantially from their theoretical values and risk premium estimates are distorted;\footnote{This is one of the major challenges to measuring variance risk premia, and \cite{AKM14} show that market rates for S\&P500 variance swaps were very often more than 5\% different from their theoretical fair value during the turbulent year surrounding the Lehman Brothers collapse in September 2008. Some recent work on the variance risk premia such as \cite{ELW10} and \cite{KS14} employs market quotes rather than theoretical values, but long time series of market quotes are not available for most alternative swap rates other than, of course, the futures price.}
\item Monitoring partition: This refers to the monitoring of the realised pay-off. The common practice for risk premium analysis is to use daily log returns (i.e. the daily partition) because this corresponds to the terms and conditions in most variance swap contracts;\footnote{There is a large literature on the use of ultra-high frequency data for realised variance when it is used as an estimator of the quadratic variation, but this is more relevant to very short-term speculators and will not be discussed in this paper.}
\item Sampling frequency: This is the frequency of the \gls{pnl} time series for moment swaps. As already noted by \cite{AKM14}, when this is less than the investment horizon it is not easy to produce a non-overlapping time series which corresponds to an investment that can actually be made. \cite{G79} proves that linear interpolation over option prices does not produce an investable return, and a trading strategy that runs until maturity and then rolls on to another swap of a similar maturity yields a risk premia time series with a systematically varying investment horizon. So to analyse the premium at a constant horizon one must construct synthetic, constant-maturity swap rates with considerable care, using linear interpolation over price changes (total returns) rather than option prices themselves.\footnote{This error in interpolation is unfortunately common in the academic literature, and when a term structure is almost always increasing (or  decreasing) with term the investable constant-maturity series exhibits very different characteristics from the non-investable one. For instance, the investable indices for \gls{vix} futures have a very strong downward trend due to the positive roll cost, especially at the short end of the term structure. See \cite{AK13} for further information. Fixed-strike European options are a case in point here: call prices always increase with term (up to about one or two years, when the discount rate effect might dominate) and put prices increase with term when the trend is negative and also -- up to about 6 months -- when the trend is positive. The exact point when put prices start to decrease with term depends on the strike and the trend in the underlying futures.}
\end{enumerate} 

\subsection{Literature Review}

We now summarise how these choices have been made in the recent model-free strand of the literature on variance risk premia. Almost all these papers study the S\&P500, with the notable exception of \cite{AB13}, who examine currency markets from 2003--2009 and \cite{TS10}, who focus on oil and gas markets from 1996--2006. The S\&P500 studies of \cite{CW09}, \cite{ELW10}, \cite{AKM14} and \cite{KS14} all employ a sample starting in 1996, the shortest (Carr and Wu) ending in 2003 and the longest (Ait-Sahalia et al.) ending in 2010. In every case the investment horizon corresponds to the maturity of the swap:\footnote{Only \cite{KS14} also allow it to be shorter.} 1 month in \cite{CW09} and \cite{TS10}; 1 and 3 months in \cite{AB13}; and the same term structure of horizons from 2 to 24 months in \cite{ELW10}, \cite{AKM14} and  \cite{KS14}.

Regarding the three criteria discussed in the previous sub-section: (1) All the cited papers define realised variance as the sum of squared log returns, so there are approximation errors in the model-free variance swap rate which are summarised in the next section; (2) The realised variance is monitored at the daily frequency.\footnote{Except that \cite{AB13} also examine intra-day monitoring.} This corresponds to industry practice in the terms and conditions of most variance swaps; (3) Two studies examine risk premia that are not actually investable (\cite{CW09} and \cite{AB13}) and all of them, with the notable exception of \cite{ELW10}, employ overlapping data, which induces artificial autocorrelation in the risk premium estimates.\footnote{\cite{CW09} use Fama-French factors to examine the determinants of the 1-month S\&P500 \gls{vrp}, finding no significance except for the excess market return and no significant asymmetric response to market shocks. However, these findings could be influenced by their data being autocorrelated and non-investable.}  

\cite{KNS13} study the 1-month S\&P500 variance and skewness risk premia during the period 1996--2012. By using the `aggregating' variance and skewness introduced by \cite{N12} they avoid the problem of errors in the model-free swap rate that besets all former studies of the variance risk premia. However, unless one assumes S\&P500 futures follow a martingale in the physical measure they use biased and relatively efficient estimators for realised variance and skewness. Using time series of about 200 monthly observations they conclude that the skewness risk premium is driven almost entirely by the \gls{vrp} 
and that strategies designed to capture skewness and hedge out exposure to variance risk earn an insignificant risk premium. They extend some results to daily observations on 1-month risk premia, but they employ linear interpolation over option replication portfolios, so their daily time series do not represent investable returns, and their data are overlapping. Also, they do not explore the common factors which might determine both variance and skewness risk premia.

\section{Theory}\label{sec:theory}
\subsection{Variance Swaps}
A conventional variance swap on the forward price $F=\mathrm e^x$ of a financial asset pays the realised sum of squared log returns over some investment horizon $[t,T]$ in exchange for a fixed swap rate, with settlement at maturity. The realised leg is monitored under some partition, typically daily. For ease of exposition but without loss of generality we employ the regular partition $\boldsymbol\Pi_n:=\left\{\left.t+\tfrac{i}{n}(T-t)\right|i=0,\ldots,n\right\}$, so the realised variance is given by\footnote{Here and in the following we ignore the averaging in swap rates and realised variance, since including this only adds a non-essential level of complexity to our notation.}
\begin{equation}\label{eq:rv}
rv_n :=\sum_{\boldsymbol\Pi_n}\hat{x}^2,
\end{equation}
where $\hat{x}$ are increments in the log forward price $x$ along  $\boldsymbol\Pi_n$. 
The  fair-value swap rate is fixed at the inception of the swap (time $t$) so that the expected pay-off on the swap under the risk-neutral measure $\mathbbm{Q}$ is zero. Following \cite{DDKZ99} and others, the variance swap rate is typically approximated as
\begin{equation}\label{eq:vsr}
\mathbbm{E}_t^\mathbbm{Q}[rv_n] \approx 2\int_{\mathbbm{R}^+}k^{-2}q(k)dk,
\end{equation}
where $q(k)$ denotes the price of a vanilla \gls{otm} option with maturity $T$ and strike $k$ and $\mathbbm{E}_t^\mathbbm{Q}[.]$ denotes the risk-neutral expectation conditional on all information available at time $t$.

Under hypothetical continuous monitoring the realised variance converges to the quadratic variation $\langle x\rangle$ and an estimator for the \gls{vrp} (or, more precisely, the quadratic variation risk premium) is the conditional expected excess return on the continuously-monitored variance swap under the physical measure $\mathbbm{P}$, i.e.
\begin{equation}\label{eq:est}
\mu^{\langle x\rangle}:= \mathbbm{E}_t^\mathbbm{P}\left[\langle x\rangle\right]-\mathbbm{E}_t^\mathbbm{Q}\left[\langle x\rangle\right]= \mathbbm{E}_t^\mathbbm{P}\left[\left(1-m_t\right)\langle x\rangle\right], 
\end{equation}
where  $\mathbbm{E}_t^\mathbbm{P}[.]$  denotes the conditional expectation under $\mathbbm{P}$, and $m_t:=\left.\tfrac{d\mathbbm{Q}}{d\mathbbm{P}}\right|_{\mathcal{F}_t}$ denotes the corresponding conditional pricing kernel.
Under the further assumption that $F$ follows a pure diffusion the approximation  \eqref{eq:vsr} becomes exact.\footnote{But  in practice options are not traded at a continuum of strikes. See \cite{JT05} for further details in the approximation errors caused by numerical integration over a finite set of strikes.} Then
\begin{equation}\label{eq:estrv}
\mu^{\langle x\rangle}= \mathbbm{E}_t^\mathbbm{P}\left[\langle x\rangle\right]- 2\int_{\mathbbm{R}^+}k^{-2}q(k)dk.
\end{equation}
To implement the swap we cannot monitor the realised leg continuously, instead we monitor it using the partition $\boldsymbol\Pi_n$. In this case we can use the following estimator for the  \gls{vrp}:
\begin{equation}\label{eq:est3}
\mu_n^{rv}:=  \mathbbm{E}_t^\mathbbm{P}\left[\left(1-m_t\right)rv_n\right].
\end{equation} 
A popular sample estimate for $\mu_n^{rv}$  is the wealth $w_n^{rv}$ that the investor realises from entering the swap. This is the profit or loss (\gls{pnl}) on the discretely-monitored swap, which, with the swap-rate approximation \eqref{eq:vsr}, is given by
\begin{equation}\label{eq:pnlconventionalvarianceswap}
w_n^{rv}:=\sum_{\boldsymbol\Pi_n}\hat{x}^2-2\int_{\mathbbm{R}^+}k^{-2}q(k)dk.
\end{equation}
This is the basis of the swaps-based approach to measuring the variance risk premium, pioneered by \cite{CW09} and since used by many other authors, as described in the previous sub-section. 

\subsection{The Aggregation Property}

Taking a completely different approach to the model-free variance swap literature,  both \cite{N12} and \cite{B14} re-define the realised variance in such a way that  there exists an exact, model-free fair-value variance swap rate (under the minimal assumption of no arbitrage). Furthermore, \cite{N12} proves that the same rate applies irrespective of the monitoring frequency of the floating leg, 
provided the realised characteristic satisfies the aggregation property (AP) stated below: 

Let $\phi:\mathbbm{R}^n\rightarrow\mathbbm{R}$  define a pay-off function on the process $\mathbf{x}$. It need not be a martingale, for instance \cite{N12} considers $\mathbf{x}=\left( \ln F, v\right)$ where $v$ is a generalised variance process on $F$.  
Then the AP may be written: 
\begin{equation}\label{eq:aggregationpropertyneuberger}
\mathbbm{E}\left[\sum_{\boldsymbol\Pi_{m}}\phi\left(\mathbf{\hat{x}}\right)\right]=\mathbbm{E}\left[\phi\left(\mathbf{x}_{_T}-\mathbf{x}_{_0}\right)\right] \quad \forall \,\,\mbox{partitions}\,\, \boldsymbol\Pi_{m},
\end{equation}
One of the main contributions of \cite{N12} is to propose a definition of the realised third moment that is computed from daily returns (and vanilla option prices) which provides an unbiased estimate of the true third moment of long-horizon returns. However, we show in section \ref{sec:bias} that this is the case only when futures prices follow a martingale process in the physical measure, i.e. in the absence of an \gls{erp}. 

\cite{N12} takes the original step of including conditional expected pay-offs in $\mathbf{x}$, allowing the floating leg of a swap to encompass information about serial dependence. He then classifies all pay-offs $\phi$ which satisfy \eqref{eq:aggregationproperty} for two specific processes: (i) $\mathbf{x}:=\left(F,v\right)^\prime$ and $v$ being the conditional variance process $v_t:=\mathbbm{E}^\mathbbm{Q}_t\left[\left(F_T-F_t\right)^2\right]$; as well as (ii) $\mathbf{x}:=\left(x,v^\eta\right)^\prime$ and $v^\eta$ being a generalised conditional process $v^\eta_t:=\mathbbm{E}^\mathbbm{Q}_t\left[\eta\left(x_T-x_t\right)\right]$ with $\lim_{\hat{x}\rightarrow0}\eta\left(\hat{x}\right)/\hat{x}^2=1$. Within the set of pay-offs pertaining to (ii), and using the specific conditional entropy variance $\eta\left(\hat{x}\right):=2\left(\hat{x}\mathrm e^{\hat{x}}-\mathrm e^{\hat{x}}+1\right)$, Neuberger identifies the pay-off
\begin{equation}\label{eq:Nskew}
\psi\left(\mathbf{\hat{x}}\right):=3\hat{v}^\eta\left(\mathrm e^{\hat{x}}-1\right)+\tilde{\psi}\left(\hat{x}\right),
\end{equation}
with $\tilde{\psi}\left(\hat{x}\right):=6\left(\hat{x}\mathrm e^{\hat{x}}-2\mathrm e^{\hat{x}}+\hat{x}+2\right)$, where $\hat{v}^\eta$ denotes an increment in $v^\eta$, and argues that it approximates the third moment of log returns as long as $F$ follows a martingale, since the first term then has expectation zero and $\lim_{\hat{x}\rightarrow 0}\tilde{\psi}\left(\hat{x}\right)/\hat{x}^3=1$.\footnote{The entropy variance process can be replicated using $v^\eta_t=2F_t^{-1}\int_{\mathbbm{R}^+}k^{-1}q(k)dk$.}

\subsection{Discretisation Invariance}
Restricting the general framework of \cite{N12}, \cite{AR16} consider only those processes $\mathbf{x}$ adapted to the pay-off $\phi$ that are deterministic functions of $\mathbbm{Q}$-martingales. For such $\mathbf{x}$ they term the contracts $\phi$ which satisfy \eqref{eq:aggregationpropertyneuberger} the class of Discretisation-Invariant (DI) swap contracts. With the further restriction that  $\mathbf{x}$ contains only $\mathbbm{Q}$-martingales and their logarithms, they characterise an entire vector space of DI $\phi$-swaps and, rather than a single definition for realised skewness as in \cite{N12} they obtain infinitely many types of higher-moment and other pay-off with aggregating characteristics, and which may be more tightly priced than standard moment swaps. With this choice for $\mathbf{x}$ the vector space also encompasses an alternative definition of the AP by \cite{B14}.

In this paper we are concerned with pay-offs $\phi$ which satisfy the \gls{ap} in the case where $\mathbf{x}$ contains only the forward prices of $n$ tradable assets in an arbitrage-free market and derive a dynamic trading strategy in these tradable assets that replicates the \gls{pnl} on the swap. Finally, we demonstrate that our general framework encompasses third and fourth moment pay-offs when we include some specific claims on the forward price of a single financial asset, each claim being replicable using portfolios of vanilla \gls{otm} options.\vspace{12pt}

\noindent {\bf Theorem 1:} The pay-offs $\phi$ that satisfy \eqref{eq:aggregationproperty} for a $\mathbbm{Q}$-martingale $\mathbf{x}$ form a vector space:
\begin{equation*}
\mathbbm{F}:=\left\{\left.\phi\left(\mathbf{\hat{x}}\right)=\boldsymbol\alpha^\prime\mathbf{\hat{x}}+\text{tr}\left(\boldsymbol\Omega\mathbf{\hat{x}}\mathbf{\hat{x}}^\prime\right)\right|\boldsymbol\alpha\in\mathbbm{R}^n,\boldsymbol\Omega^\prime=\boldsymbol\Omega\in\mathbbm{R}^{n\times n}\right\}.
\end{equation*}
Those pay-offs associated with $\boldsymbol\alpha$ represent linear strategies of buying and holding a portfolio of claims and can be seen as degenerated swaps. In the following we will focus on pay-offs associated with $\boldsymbol\Omega$, which represent the products of price changes in pairs of claims. In essence, these pay-offs measure serial dependence.\vspace{12pt}

%

\noindent {\bf Theorem 2:} When $\phi\in\mathbbm{F}$ the \gls{ap} yields the model-free fair-value swap rate
\begin{equation*}
s_t^{\phi}=\text{tr}\left(\boldsymbol\Omega\left[\boldsymbol\Sigma_t-\mathbf{x}_t\mathbf{x}_t^\prime\right]\right),
\end{equation*}
where $\boldsymbol{\Sigma}_t:=\mathbbm{E}^\mathbbm{Q}_t\left[\mathbf{x}_T\mathbf{x}_T^\prime\right]$. The swap can then be replicated in discrete time by means of a static trading strategy in $\boldsymbol{\Sigma}$ and a dynamic trading strategy in $\mathbf{x}$ according to
\begin{equation*}
\hat{\pi}_n^\phi=\boldsymbol\alpha^\prime\mathbf{\hat{x}}+\text{tr}\left(\boldsymbol\Omega\left[\boldsymbol{\hat{\Sigma}}-2\mathbf{x}_{t-}\mathbf{\hat{x}}^\prime\right]\right),
\end{equation*}
where $\boldsymbol{\hat{\Sigma}}$ denotes an increment in $\boldsymbol{\Sigma}$ along the monitoring partition $\boldsymbol\Pi_n$ and $\mathbf{x}_{t-}$ denotes the value of $\mathbf{x}$ prior to the increment $\mathbf{\hat{x}}$. We refer to this as an aggregating $\phi$-swap. Further
\begin{equation*}
\mathbbm{E}_{t-}^\mathbbm{P}\left[\hat{\pi}_n^\phi\right]=\mathbbm{E}_{t-}^\mathbbm{P}\left[\left(1-m_{t-}\right)\phi\left(\mathbf{\hat{x}}\right)\right]+\xi^\phi_{t-}
\end{equation*}
where $\xi^\phi_{t-}:=\mathbbm{E}_{t-}^\mathbbm{P}\left[\hat{m}_t\phi\left(\mathbf{x}_{_T}-\mathbf{x}_t\right)\right]$, $t-$ is the monitoring time prior to $\mathbf{\hat{x}}$ and $\hat{m}_t=m_t-m_{t-}$.\vspace{12pt}

\noindent Theorem 2 explains how to replicate (or sample) the incremental P\&L that accrues to the issuer of an aggregating $\phi$-swap who pays the fixed swap rate $\mathbbm{E}^\mathbbm{Q}\left[\phi\left(\mathbf{x}_T-\mathbf{x}_t\right)\right]$ and receives the floating pay-off $\sum_{\boldsymbol\Pi_n}\phi\left(\mathbf{\hat{x}}\right)$ along the monitoring partition. It also shows that the expected \gls{pnl} on an aggregating swap is the sum of the risk premium paid for the realised pay-off $\phi\left(\mathbf{\hat{x}}\right)$ and a noise term $\xi^\phi$.

We now focus on the case where the tail risk premia are derived from DI moment swaps on power log contracts.\footnote{Early results on power contracts alone yields very similar empirical results and are therefore not presented here for brevity.} Denote by $X_t^{(p)}:=\mathbbm{E}_t^\mathbbm{Q}\left[x_T^p\right]$ the fair-value forward price process of the $p$-th power log contract on $F$. 
According to the replication theorem of \cite{CM01}, this expectation can be expressed as
\begin{equation}\label{eq:powerlogcontracts}
X_t^{(p)}=x_t^p+\int_{\mathbbm{R}^+}\gamma_p(k)q(k)dk,
\end{equation}
where $\gamma_1(k):=k^{-2}$ and $\gamma_p(k):=p(\ln k)^{p-2}k^{-2}\left(n-1-\ln k\right)$ for $n\ge 2$.\footnote{We may also consider the alternative replication scheme:
	\begin{equation*}
	X_t^{(p)}=x_0^p+px_0^{p-1}\left(\tfrac{F_t}{F_0}-1\right)+\int_0^{F_0}\gamma_p(k)P(k)dk+\int_{F_0}^\infty\gamma_p(k)C(k)dk,
	\end{equation*}
	where $P(k)$ and $C(k)$ denote the time-$t$ forward prices of a vanilla put and call with strike $k$ and maturity $T$, respectively. This representation involves options that are \gls{otm} at time zero and therefore describes simple buy-and-hold strategies. By contrast, the portfolio in \eqref{eq:powerlogcontracts} is based on options that are \gls{otm} at time $t$, and due to the stochastic separation strike $F_t$ continuous rebalancing between puts and calls is necessary for replication. From a theoretical perspective the two representations are exchangeable. However, \eqref{eq:powerlogcontracts} may be favourable for computing the fair-value swap rate, since \gls{otm} options are more liquidly traded, while the alternative scheme may be preferrable for replication.} Now we give some examples of DI moments swaps based on these power log contracts.\vspace{12pt}


\noindent {\bf Variance Swap:} Consider the pay-off $\phi\left(\mathbf{\hat{x}}\right)=\hat{X}^2$ corresponding to $\mathbf{x}=X$ and $\boldsymbol\Omega=1$. By Theorem 1 the fair-value swap rate is $s^\phi_{_0}=s^{(2)}_{_0}:=\mathbbm{E}_t^\mathbbm{Q}\left[\left(x_T-X_t\right)^2\right]$, the second moment of the distribution of $x_T$, 
and according to Theorem 2 the P\&L yields
$
\hat{\pi}_t^{(2)}:=\hat{X}_t^{(2)}-2X_{t-1}\hat{X}_t
$. 
Hence, this swap can be hedged by selling a squared log contract and dynamically holding $2X_{t-1}$ log contracts from time $t-1$ to $t$. We can observe empirically that the P\&L on this variance swap is almost perfectly correlated with that on Neuberger's variance swap.\vspace{12pt}

\noindent {\bf Third-Moment Swap:} Let $\mathbf{x}=\left(X,X^{(2)}\right)^\prime$ and consider the pay-off corresponding to
\begin{equation*}
\boldsymbol\Omega=\left[
\begin{array}{cc}
-2X_{_0}&\tfrac{1}{2}\\
\tfrac{1}{2}&0
\end{array}\right].
\end{equation*}
The fair-value swap rate is $s^\phi_{_0}=s^{(3)}_{_0}:=\mathbbm{E}_t^\mathbbm{Q}\left[\left(x_T-X_t\right)^3\right]$ and 
according to Theorem 2 we have 
$
\hat{\pi}_t^{(3)}:=\hat{X}_t^{(3)}-h_{2t}^{(3)}\hat{X}_t^{(2)}-h_{1t}^{(3)}\hat{X}_t
$ 
with $h_{2t}^{(3)}:=2X_{_0}+X_{t-1}$ and $h_{1t}^{(3)}:=X_{t-1}^{(2)}-4X_{_0}X_{t-1}$. Hence, the swap can be hedged by selling a cubed log contract and dynamically holding $h_{2t}^{(3)}$ squared log contracts as well as $h_{1t}^{(3)}$ log contracts from time $t-1$ to $t$.\vspace{12pt}

\noindent {\bf Fourth-Moment Swap:} Let $\mathbf{x}=\left(X,X^{(2)},X^{(3)}\right)^\prime$ and consider the pay-off corresponding to
\begin{equation*}
\boldsymbol\Omega=\left[
\begin{array}{ccc}
3X_{_0}^2&-\tfrac{3}{2}X_{_0}&\tfrac{1}{2}\\
-\tfrac{3}{2}X_{_0}&0&0\\
\tfrac{1}{2}&0&0
\end{array}\right].
\end{equation*}
Then $s^\phi_{_0}=s^{(4)}_{_0}:=\mathbbm{E}_t^\mathbbm{Q}\left[\left(x_T-X_t\right)^4\right]$ 
and  
$
\hat{\pi}_t^{(4)}:=\hat{X}_t^{(4)}-h_{3t}^{(4)}\hat{X}_t^{(3)}-h_{2t}^{(4)}\hat{X}_t^{(2)}-h_{1t}^{(4)}\hat{X}_t
$ 
with $h_{3t}^{(4)}:=3X_{_0}+X_{t-1}$, $h_{2t}^{(4)}:=-3X_{_0}^2-3X_{_0}X_{t-1}$ and $h_{1t}^{(4)}:=X_{t-1}^{(3)}-3X_{_0}X_{t-1}^{(2)}+6X_{_0}^2X_{t-1}$ and the swap can be hedged by selling a quartic log contract and  holding $h_{3t}^{(4)}$ cubed log contracts, $h_{2t}^{(4)}$ squared log contracts and $h_{1t}^{(4)}$ log contracts from  $t-1$ to $t$.\vspace{12pt}

\section{Long-Term Tail Risk Premia}\label{sec:LTTRP}

Contrary to \cite{KNS13}, we believe that the upper and lower tails of the long-term log return distribution do capture relatively uncorrelated risk premia, and our objective is to reveal these by applying the aggregating moment swaps developed in the previous section. Besides the choice of the realised characteristic, the methodology used for constructing the estimators is crucial. This section therefore discusses the bias and efficiency of different moment risk premium estimators. Proofs are in the Appendix.

\subsection{Bias}\label{sec:bias}

We define the expectation risk premium (ERP) as the compensation required for exposure to a forward price $F$ of a risky asset, or in our case its logarithm, at time $T>t$, denoted $x_T$. It is the difference between its expected value under the physical and risk-neutral measures, $\mathbbm{P}$ and $\mathbbm{Q}$. So an estimator for the ERP is $\mathbbm{E}_t^\mathbbm{P}[x_T]-\mathbbm{E}_t^\mathbbm{Q}[x_T]$. Assuming no-arbitrage so the price of the contract is a martingale in $\mathbbm{Q}$, then the log price $x_t$ is not a martingale. However the log contract which has time $t$ value $X_t=\mathbbm{E}_t^\mathbbm{P}[x_T]$, is a martingale. Using the log contract value $X_t$, and noting that $X_T=x_T$ the ERP may be written  $\mathbbm{E}_t^\mathbbm{P}[X_T]-X_t$. So the ERP is just the expected P\&L under $\mathbbm{P}$ on the log contract. 

The VRP (DRP) is the compensation required for exposure to the variance (dispersion) of $x_T$, as perceived today (time $t$), about it's fair value at time $t$. In the swap-based approach the VRP is identified with the P\&L on a delta-hedged short log contract, i.e. $$ \sum_{i=t}^{T-1} F_{i}^{-1}\left[\mathbbm{E}_t^\mathbbm{P}[F_{i+1}]-F_{i} \right] - \left[\mathbbm{E}_t^\mathbbm{P}[X_T]-X_t \right]  $$
Since $x_T=X_T$ we may equivalently consider the dispersion of the value of the log contract at $T$. But what centre should the dispersion be based around,  $\mathbbm{E}_t^\mathbbm{P}[X_T]$ or  $\mathbbm{E}_t^\mathbbm{Q}[X_T]=X_t$? Subtracting the second moment of $X_T$ under $\mathbbm{Q}$ from its second moment under $\mathbbm{P}$ has the effect of including the ERP in the VRP measure, because the two second moments have different centres, i.e. $\mathbbm{E}_t^\mathbbm{Q}[x_T]$ and $\mathbbm{E}_t^\mathbbm{P}[x_T]$ respectively. Centering on $\mathbbm{E}_t^\mathbbm{P}[X_T]$ is correct if we want to disengage the ERP and the VRP, so from the current value $X_t$ first we move to $\mathbbm{E}_t^\mathbbm{P}[X_T]$ with the ERP and then we measure dispersion about $\mathbbm{E}_t^\mathbbm{P}[X_T]$. The alternative, i.e. to measure dispersion about $X_t$, i.e. about $\mathbbm{E}_t^\mathbbm{Q}[X_T]$ would give a single risk premium estimator which includes the ERP [insert graph to depict this].

However, with the definition \eqref{eq:rv} for the realised variance, the integral in the second term of the r.h.s. of \eqref{eq:pnlconventionalvarianceswap} is not the exact fair value of $rv_n$ under $\mathbbm{Q}$. The use of discrete monitoring, and possible jumps in $F$ induce a bias to the estimates \eqref{eq:est3}. That is, because $\mathbbm{E}_t^\mathbbm{Q}[w_n^{rv}] =: \varepsilon_n^{rv} \ne 0$, 
the estimator \eqref{eq:est3} for the \gls{vrp} may be written  $ \mu_n^{rv}=\mathbbm{E}_t^\mathbbm{P}[w_n^{rv}-\varepsilon_n^{rv}] $. In fact, the previous literature has estimated $\mathbbm{E}_t^\mathbbm{P}[w_n^{rv}]$, which is not $\mu_n^{rv}$ but  $\tilde{\mu}_n^{rv}= \mu_n^{rv}+ \varepsilon_n^{rv}$.

To avoid this bias, \cite{N12} proposes to replace the realised characteristic \eqref{eq:rv} with an alternative dispersion estimator called the log variance, defined as 
\begin{equation}\label{eq:lv}
lv_n :=\sum_{\boldsymbol\Pi_n}2\left(\mathrm e^{\hat{x}}-1-\hat{x}\right).
\end{equation}
 If \eqref{eq:lv} is used in place of \eqref{eq:rv} for the floating leg of a variance swap then the fair-value swap rate is exact, provided only that $F$ is a $\mathbbm{Q}$-martingale, which is always the case in the absence of arbitrage. In fact, for any monitoring partition, and even with jumps in $F$, we have:
 \begin{equation}\label{eq:vsrn}
  \mathbbm{E}_t^\mathbbm{P}[lv_n] = 2\int_{\mathbbm{R}^+}k^{-2}q(k)dk.
  \end{equation} 
The key to this property is that \eqref{eq:lv} satisfies the AP in $\mathbbm{Q}$. In other words, because $F$ is a martingale in $\mathbbm{Q}$ we have:
  \begin{equation}\label{eq:ap}
  \mathbbm{E}_t^\mathbbm{Q}\left[\sum_{\boldsymbol\Pi_n}2\left(\mathrm e^{\hat{x}}-1-\hat{x}\right)-2\left(\mathrm e^{(x_{_T}-x_{t})}-1-(x_{_T}-x_{t})\right)\right]=0
  \end{equation}
  Now we can use the following estimator for the \gls{vrp}:
   \begin{equation}\label{eq:estlv}
 \mu^{lv}:=  \mathbbm{E}_t^\mathbbm{P}\left[\left(1-m_t\right)lv\right]  ,
   \end{equation}
   and an estimate for  $\mu^{lv}$ is the \gls{pnl} realised on the log variance swap, i.e. 
\begin{equation}
w_n^{lv}:=\sum_{\boldsymbol\Pi_n}2\left(\mathrm e^{\hat{x}}-1-\hat{x}\right)-2\int_{\mathbbm{R}^+}k^{-2}q(k)dk.
\end{equation}
Because \eqref{eq:vsrn} is exact, this log variance swap has zero $\mathbbm{Q}$-bias, i.e. $\varepsilon_n^{lv}:= \mathbbm{E}_t^\mathbbm{Q}[w_n^{lv}]=0$. 
If $F$ were also a martingale in $\mathbbm{P}$, then \eqref{eq:ap} would also hold under $\mathbbm{P}$. However, if the equity risk premium is not zero then $F$ is not a $\mathbbm{P}$-martingale. As a result, the $\mathbbm{P}$-bias is \textit{not} zero, i.e. $\eta_n^{lv}:=\mathbbm{E}_t^\mathbbm{P}[w_n^{lv}] - \mu_n^{lv}\ne0$.  

Our estimator for the VRP $\mu^{(2)}$ is the expected \gls{pnl} on a swap that pays the sum of squared price changes in the log contract. Such a design combines the association with the second moment from the conventional variance swap with the \gls{ap} from Neuberger's alternative definition. The \gls{pnl} of our second moment swap is
\begin{equation}\label{eq:pnlsecondmomentswap}
\pi_n^{(2)}:=\sum_{\boldsymbol\Pi_n}\hat{X}^2-\mathbbm{E}_t^\mathbbm{Q}\left[\left(X_T-X_t\right)^2\right]
\end{equation}
at maturity, where $\hat{X}$ are increments in $X$, and both biases, $\varepsilon_n^{(2)}$ and $b_1^{(2)}$, are zero by construction. But, as when using Neuberger's log variance, is it only when the forward price is also a martingale under $\mathbbm{P}$, i.e. in the absence of an \gls{erp}, that the realised second moment pay-off also aggregates under the physical measure so that  $\pi_n^{(2)}$ is an unbiased estimator for the second moment risk premium.

Now we introduce more general notation and consider the forward prices $\mathbf{x}$ of $n$ tradable assets, which under the assumption of no arbitrage follow a multivariate $\mathbbm{Q}$-martingale over the investment horizon $[t,T]$. For some pay-off $\phi:\mathbbm{R}^n\rightarrow\mathbbm{R}$ the \gls{pnl} on a $\phi$-swap wrt $\mathbf{x}$ is
\begin{equation}\label{eq:phiswap}
\pi_n^{\phi}:=\sum_{\boldsymbol\Pi_n}\phi\left(\mathbf{\hat{x}}\right)-s_t^{\phi},
\end{equation}
where $s_t^{\phi}$ denotes the fair-value $\phi$-swap rate at inception and $\mathbf{\hat{x}}$ denote increments in $\mathbf{x}$.

The absence of a $\mathbbm{Q}$-bias for a $\phi$-swap, i.e. requiring $\varepsilon_n^{\phi}=0$ and $\varepsilon_1^{\phi}=0$ in particular is equivalent to Neuberger's \gls{ap}, which holds for the pay-off $\phi$ and a price process $\mathbf{x}$ under the risk-neutral measure iff
\begin{equation}\label{eq:aggregationproperty}
\mathbbm{E}_t^\mathbbm{Q}\left[\sum_{\boldsymbol\Pi_n}\phi\left(\mathbf{\hat{x}}\right)-\phi\left(\mathbf{x}_T-\mathbf{x}_t\right)\right]=0
\end{equation}
for all partitions $\boldsymbol\Pi_n$, where $s_t^{\phi}:=\mathbbm{E}_t^\mathbbm{Q}\left[\sum_{\boldsymbol\Pi_n}\phi\left(\mathbf{\hat{x}}\right)\right]$ is the fair-value swap rate.\footnote{A variation of the \gls{ap}, namely $$\mathbbm{E}^\mathbbm{Q}\left[\sum_{i=1}^m\tilde\phi\left(F_{t_i},F_{t_{i-1}}\right)\right]=\mathbbm{E}^\mathbbm{Q}\left[\tilde\phi\left(F_T,F_0\right)\right],$$ is discussed in \cite{B14}. This property is less general than the \gls{ap} in that only a univariate process is considered, but more general in that the function need not be defined on increments only.} The analogy with \eqref{eq:pnlsecondmomentswap} is obvious, and it is easy to see that the \gls{ap} {\it does not} hold for $\phi\left(\hat{x}\right)=\hat{x}^2$, the conventional variance pay-off. In fact, the \gls{ap} does not hold for any $\phi\left(\hat{x}\right)=\hat{x}^n$ with $n\ge 2$, which substantiates our initial statement that long-term and short-term tail risk are two very different phenomena. Cubed, quartic or higher-power short-term returns are essentially a measure of (model-dependent) jump risk and largely irrelevant for the moments of the long-term distribution.

Yet, if the \gls{ap} holds the expected realised pay-off is path-independent, and even if investors differ in their views about jump risk in an incomplete market they will still agree on the fair-value swap rate $s_t^{\phi}=\mathbbm{E}_t^\mathbbm{Q}\left[\phi\left(\mathbf{x}_T-\mathbf{x}_t\right)\right]$, as long as they agree on the market-implied measure at time $T$.\footnote{Furthermore, the swap rate can be expressed in terms of vanilla \gls{otm} options by applying the replication theorem of \cite{CM01}.}
Then the \gls{pnl} on a $\phi$-swap is an estimator for the $\phi$-risk premium
\begin{equation*}\label{eq:phiriskpremium}
\mu^{\phi}:=\mathbbm{E}_t^\mathbbm{P}\left[\left(1-m_t\right)\phi\left(\mathbf{x}_T-\mathbf{x}_t\right)\right],
\end{equation*}
and the $\mathbbm{P}$-bias $b_n^{\phi}:=\mathbbm{E}_t^\mathbbm{P}\left[\pi_n^\phi\right]-\mu^\phi$ corresponds to the amount by which the \gls{ap} is violated under the physical measure.

However, unlike in the case of the log variance, $\lambda$, the association of the pay-off \eqref{eq:Nskew} with the third moment depends on the martingale assumption and  $F$ does not follow a martingale in the presence of an \gls{erp}.\footnote{Ex-post observations on index futures often exhibit pronounced trends, as is the case for the S\&P 500, and therefore the martingale assumption under the physical measure is not realistic.} In other words, $\varepsilon_n^\psi=0$ but generally $b_n^\psi\ne 0$ and $\psi$ does therefore {\it not} represent a suitable alternative to $\hat{x}^3$ for measuring short-term skewness. That is, Neuberger's realised skewness is biased under the physical measure, and the bias is caused by the \gls{erp}.\footnote{Through simulation, \cite{N12} shows that his third moment estimator is much more efficient than the standard third moment, which together with the $\mathbbm{P}$-bias yields an estimator that is very likely to be wrong. He discusses the issue in Proposition 7. This may explain why \cite{KNS13} (calculating their skewness risk premium based on $\psi$) their variance and skewness risk premia are highly correlated, being related via the leverage effect (see e.g. \cite{B96}).}

If the \gls{ap} holds, i.e. $\mathbbm{E}^\mathbbm{Q}\left[\left.\pi_n^\phi\right|\mathcal{F}_t\right]=0$, the \gls{pnl} \eqref{eq:phiswap} is an unbiased estimator for the realised $\phi$-risk premium
\begin{equation*}
\mu_n^\phi:=\mathbbm{E}^\mathbbm{P}\left[\left(1-m_t\right)\left.\sum_{\boldsymbol\Pi_n}\phi\left(\mathbf{\hat{x}}\right)\right|\mathcal{F}_t\right],
\end{equation*}
with the uncertainty about this estimator being
\begin{equation*}
\left(\sigma_n^\phi\right)^2:=\mathbbm{E}^\mathbbm{P}\left[\left.\left(\sum_{\boldsymbol\Pi_n}\phi\left(\mathbf{\hat{x}}\right)\right)^2\right|\mathcal{F}_t\right]-\mathbbm{E}^\mathbbm{P}\left[\left.\sum_{\boldsymbol\Pi_n}\phi\left(\mathbf{\hat{x}}\right)\right|\mathcal{F}_t\right]^2.
\end{equation*}

\subsection{Efficiency}

The focus of our research is on long-term tail risk, and the most suitable criterion for the association of a swap with the $n$-th (central) moment of the long-term return distribution is to define the swap rate as:
\begin{equation}\label{eq:momentcondition}
s^\phi_{_0}=\mathbbm{E}^\mathbbm{Q}\left[\left(x_{_T}-\mathbbm{E}^\mathbbm{Q}\left[x_{_T}\right]\right)^n\right].
\end{equation}
Again, $\psi$ is not suitable (even under the risk-neutral measure) because the Taylor expansion of $\psi$ contains higher order terms. One of our theoretical contributions is to fix this problem.

The fair-value of the pay-off $\phi\left(x_T-X_t\right)$ at time $t$ is $\mathbbm{E}^\mathbbm{Q}\left[\left.\phi\left(x_T-X_t\right)\right|\mathcal{F}_t\right]$, and this value is implied by the prices of vanilla options with maturity $T$. However, investors are also interested in the conditional true pay-off under the physical measure $\mathbbm{P}$, i.e. $\mathbbm{E}^\mathbbm{P}\left[\left.\phi\left(x_T-X_t\right)\right|\mathcal{F}_t\right]$. These two expectations generally differ if investors are not risk-neutral, and we quantify the risk premium as
\begin{equation*}
\mu^\phi:=\mathbbm{E}^\mathbbm{P}\left[\left(1-m_t\right)\left.\phi\left(x_T-X_t\right)\right|\mathcal{F}_t\right],
\end{equation*}
For a fixed investment horizon there is just one observation to be made for the \gls{pnl} on a $\phi$-forward contract:
\begin{equation*}
\pi^\phi:=\phi\left(x_T-X_t\right)-\mathbbm{E}^\mathbbm{Q}\left[\left.\phi\left(x_T-X_t\right)\right|\mathcal{F}_t\right].
\end{equation*}
This estimator for the $\phi$-risk premium is clearly unbiased since $\mathbbm{E}^\mathbbm{P}\left[\left.\pi^\phi\right|\mathcal{F}_t\right]=\mu^\phi$, but it is also very inefficient since $$\left(\sigma^\phi\right)^2:=\mathbbm{V}^\mathbbm{P}\left[\left.\pi_{_{t,T}}^\phi\right|\mathcal{F}_t\right]=\mathbbm{E}^\mathbbm{P}\left[\left.\phi\left(x_T-X_t\right)^2\right|\mathcal{F}_t\right]-\mathbbm{E}^\mathbbm{P}\left[\left.\phi\left(x_T-X_t\right)\right|\mathcal{F}_t\right]^2$$ can be very large. This is not surprising because most information contained in the price trajectory is neglected.

In the (trivial) case $\phi\left(X\right)=X$ the swap rate is zero and the uncertainty about the estimator $\pi^{(1)}=x_T-X_t$ for the \gls{erp}, $\mu^{(1)}=\mathbbm{E}^\mathbbm{P}\left[\left.x_T\right|\mathcal{F}_t\right]-X_t$, yields $\left(\sigma^{(1)}\right)^2=\mathbbm{V}^\mathbbm{P}\left[\left.\pi^{(1)}\right|\mathcal{F}_t\right]=\mathbbm{E}^\mathbbm{P}\left[\left.\left(x_T-X_t\right)^2\right|\mathcal{F}_t\right]$. This definition coincides with the integrated \gls{erp} in \cite{BGL13} but differs from \cite{BT11} who use $\mathrm e^{x_{_T}-x_t}-1$ as their estimator.

\section{Determinants of S\&P500 Long-Term Tail Risk Premia}\label{sec:empirical}

This section analyses the risk premia on S\&P500 $\phi$-swaps over an 18-year period from January 1996 to December 2013 using 30-day constant maturity P\&L time series for swaps based on Discretisation-Invariant (DI) higher-moment characteristics. Our main purpose is to investigate the common factors influencing variance and higher-moment risk premia that are relevant for long-term equity investors. In contrast to most previous studies, with the notable exception of \cite{KNS13}, we examine the risk premia based on DI realised characteristics. This is because we can derive unbiased estimates of DI risk premia from their fair-value swap rates, i.e. we do not need to rely on  market quotes which are anyway not currently available. We find empirical features in these DI  risk premia which depend on their monitoring frequency, unlike their fair-value swap rates.

Most previous studies distinguish the sampling frequency of the data from the monitoring frequency of the realised characteristic, typically employing monthly or weekly data on a daily-monitored characteristic.\footnote{For instance, \cite{KNS13} uses monthly data on daily-monitored skew swaps (and also some daily data but non-investable) and \cite{ELW10} use weekly data on daily-monitored variance swaps.} By contrast, we construct our data to match the sampling and monitoring frequencies, using daily data on daily-monitored characteristics, weekly data on weekly-monitored characteristics and monthly data on monthly-monitored characteristics (assuming 5 trading days per week and 20 trading days per month). This way, we can make inference on the properties of risk premia that are relevant for investors who monitor and rebalance positions every few days (e.g. hedge funds) as well as mutual fund and large institutional investors that typically have longer-term investment horizons. Most previous work concerns the second category of investor, but here we are also  interested in the potential benefits of short-term diversification and the immediate response of risk premia to market shocks that one can only investigate using daily (or higher frequency) data. 

To measure the ex-ante \gls{vrp}, most authors focus on the total P\&L on the swap $\pi_n^\phi$ as a realisation, which corresponds to setting the sampling frequency $\boldsymbol\Pi_s$ equal to the investment horizon $[t,T]$, and analyse it over consecutive (and possibly overlapping) sampling intervals. Our empirical findings suggest that more granular information about the risk premium can be extracted from marking the swap to market every time it is monitored. This approach increases the sample size and further enables the analysis of risk premia that relate to a constant time-to-maturity, as opposed to the periodically decreasing time-to-maturity implicit in the standard procedure. In accordance with \cite{G79}, we construct investable constant time-to-maturity time series. We present results for daily, weekly and monthly monitored characteristics  with 30 days to maturity.\footnote{Results using the same data at 90 and 180 days to maturity are presented in \cite{AR16}. By varying the maturity they infer some interesting stylised facts about the term structure of implied moment characteristics.} 

Similar to the standardisation of the third-moment swap in \cite{KNS13}, we standardise an $n$-th moment swap by dividing the change in both realised and implied by the corresponding power of the implied variance of the log-return, i.e.
\begin{equation}\label{eq:standardisedmomentswap}
\pi_t^{(\bar{n})}=\pi_t^{(n)}\left(X_{_0}^{(2)}-X_{_0}^2\right)^{-n/2}.
\end{equation}
In particular we define a skewness and a kurtosis swap on the log-return distribution by setting $\pi_t^{(\bar{3})}=\pi_t^{(3)}\left(X_{_0}^{(2)}-X_{_0}^2\right)^{-3/2}$ and $\pi_t^{(\bar{4})}=\pi_t^{(4)}\left(X_{_0}^{(2)}-X_{_0}^2\right)^{-2}$.

\subsection{Data}\label{subsec:data}
Following \cite{CW09}, \cite{T10} and others we generate observations on risk premia as the difference between the observed  realised characteristic under the physical measure and its synthetic fair value under the risk-neutral measure. As previously mentioned, much previous research on the empirical behaviour of the \gls{vrp} has used synthetic rates which yield biased estimates. 
We obtain daily closing prices $P_t$ and $C_t$ of all traded European put and call options on the S\&P500 between January 1996 and December 2013 and eliminate quotes that fulfil any of the following criteria: less than seven calendar days to maturity, more than 365 calendar days to maturity, zero trading volume, mid-price $\le 0.5$ or an implied Black Scholes volatility $\le 1\%$ or $\ge 1$. For each trading day, we further delete all quotes that refer to the same maturity if less than three different strikes are traded. 
The forward price is backed out via put-call-parity for each maturity from the pair of quotes whose strike minimises $\left|P_t-C_t\right|$. This forward price is also used as the separation strike between OTM put and call options, i.e. we use the put price for $k<S_t$ and the call price for $k\ge S_t$. 

In order to preclude static arbitrage between strikes of the same maturity, and between options of different maturities, we apply the cubic spline interpolation algorithm developed by \cite{F09}. For each day spanned by our sample this interpolation produces an equally distributed grid of OTM option prices with 2000 different strikes for each expiry date.\footnote{The strikes are equally distributed across a six-$\sigma$-range around the forward price, $\sigma$ being the average implied volatility on that day, at a given maturity. Outside the domain of the spline we assume the implied volatility is constant and equal to the implied volatility at the closest strike.} These data are then integrated numerically wrt $k$ to derive time series of daily prices \eqref{eq:powerlogcontracts} for the power log contracts, $n = 1, ... 4$. For example, the log contract is approximated by $X_t\approx x_t-\sum_{j=2}^{2000}k_j^{-2}{q}_t\left(k_j\right)\left(k_j-k_{j-1}\right)$ and similar approximations apply for $X^{(n)}_t$. 
%
Besides the daily partition $\boldsymbol\Pi_{_D}$, we include increments along the partitions $\boldsymbol\Pi_{_W}$ and  $\boldsymbol\Pi_n$, reflecting swaps that are monitored on a weekly and monthly basis. This way the observations on risk premia have the same frequency as the monitoring of the swap.

Alternative methodologies for constructing a synthetic time series of risk premia over the entire 18-year sample period include: (a) hold a swap until just before maturity and then roll over to another swap with the same initial maturity, tracking observations on its realised characteristic and swap rate; (b) linearly interpolate synthetic constant-maturity swap rates and calculate the corresponding realised characteristic on every monitoring period; or (c) hold a swap for one monitoring period, then roll over to another swap with the same initial maturity.\footnote{\cite{KNS13} (p.2184) follow (a), stating that "Our empirical analysis concentrates on trading strategies that run for a month, from the first trading day after one option expires to the next month's expiration date." \cite{CW09} (p.1319) choose the construction method (b): "At each date $t$, we interpolate the synthetic variance swap rates at the two maturities to obtain the variance swap rate at a fixed 30-day horizon. [...] Corresponding to each 30-day variance swap rate, we also compute the annualized 30-day realized variance [...]."} The risk premia obtained using method (a) have a systematically varying maturity.  Method (b) is good when the data frequency matches the maturity of the characteristic, but autocorrelation appears as an unwanted artefact when time series of higher frequencies are constructed. We use method (c) because it best facilitates an investigation of the relationship between risk premia, monitoring frequency and maturity. Because linear interpolation between prices  produces synthetic constant-maturity contracts which are not truly reflective of investable returns, it is necessary here to apply linear interpolation to the daily, weekly or monthly value increments between the two adjacent traded maturities, as proved by \cite{G79}.\footnote{Thus, the change in price from time $t-1$ to time $t$ of a contract $\Phi$ with constant time-to-maturity $\tau$ is $\hat{\Phi}_t:=\left(T_u-T_l\right)^{-1}\left[\left(T_u-t-\tau\right)\hat{\Phi}^l_t-\left(T_l-t-\tau\right)\hat{\Phi}^u_t\right]$ where $\hat{\Phi}^l_t$ and $\hat{\Phi}^u_t$ denote the increments in the prices of the contracts with fixed maturity dates $T_l$ and $T_u$. Note that increments refer now to daily, weekly or monthly increments in the constant-maturity time series, rather than the fixed-maturity series that we have used for developing the theory.} 

\subsection{Risk Premia on Moment Swaps}\label{subsec:momentriskpremia}

\begin{figure}[h!]
 \vspace{10pt}
 \begin{centering}
 \includegraphics[width=1\linewidth,trim=3.65cm 0cm 2.4cm 0cm,clip=true]{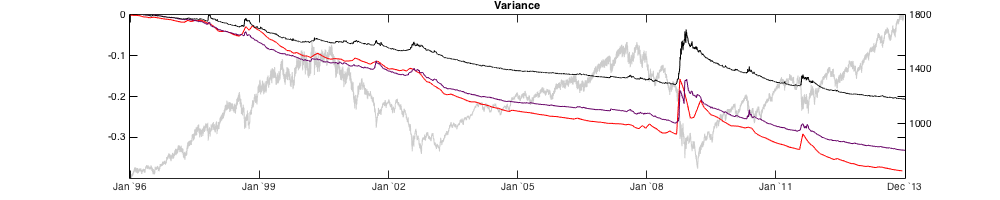}
  \includegraphics[width=1\linewidth,trim=3.65cm 0cm 2.4cm 0cm,clip=true]{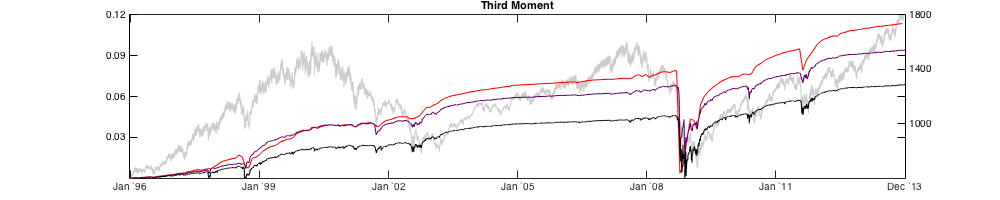}
   \includegraphics[width=1\linewidth,trim=3.65cm 0cm 2.4cm 0cm,clip=true]{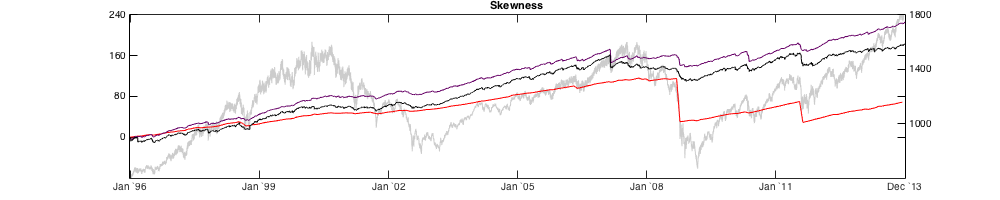}
    \includegraphics[width=1\linewidth,trim=3.65cm 0cm 2.4cm 0cm,clip=true]{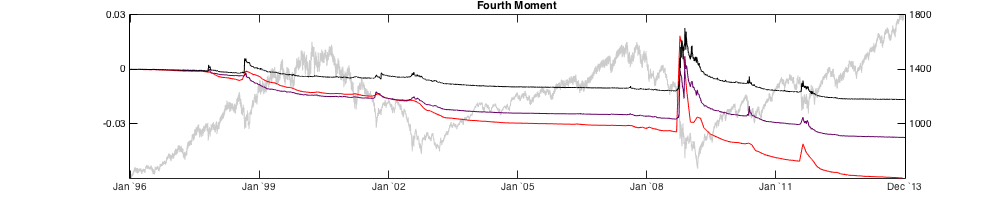}
     \includegraphics[width=1\linewidth,trim=3.65cm 0cm 2.4cm 0cm,clip=true]{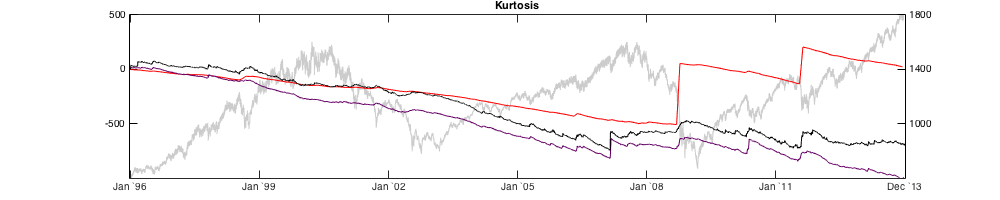}
 \par\end{centering}
 { \caption[Risk premium frequency time series]{\small Time series of cumulative 30-day  variance,  third-moment and fourth-moment risk premia 
 based on daily (black), weekly (purple) and monthly (red) monitoring. The secondary axis on the right refers to the 30-day forward contract plotted in grey.}\label{fig:varianceswapgraphsfrequency}}
 \end{figure}

\noindent Figure \ref{fig:varianceswapgraphsfrequency} depicts the cumulative risk premia for 30-day constant-maturity variance, third moment, skewness, fourth moment and kurtosis swaps over the entire sample period. 
We use a black line for daily, purple for weekly and red for monthly monitoring. 
The negative trends in both variance and kurtosis swap P\&L corresponds to the typically negative VRP already documented by many authors, and the positive association between these moments. The jumps are dominated by the realised component of the P\&L.  
 By contrast with the variance and kurtosis risk premia, the skewness premium is usually positive, but falls sharply during crisis periods when the negative skew in realised returns on equities becomes especially pronounced. This is driven by the large jump down in the realised component during September 2008 and August 2011, at the onset of the banking and Euro debt crises, respectively.  

\subsection{Diversification of Risk Premia}\label{subsec:diriskpremia}

\noindent 
 How diverse are the risk premia obtainable through trading DI moment characteristics? Table \ref{tab:swapcorrelations} presents correlations between the daily (top panel), weekly (mid panel) and monthly (bottom panel) monitored increments in  the moment risk premia depicted in Figure 1. 
 \begin{table}[h!]\small
 	\vspace{10pt}
 	\begin{center}
 		\begin{tabular}{l|c:ccccc}
 			$\boldsymbol\Pi_{_D}$ & $X$ & $\pi^{(2)}$ & $\pi^{(3)}$ & $\pi^{(\bar{3})}$ & $\pi^{(4)}$ & $\pi^{(\bar{4})}$  \\\hline
 			$F$ & 0.98 & -0.61 & 0.60 & 0.74 & -0.45 & -0.52  \\
 			$X$ & 1 & -0.66 & 0.69 & 0.71 & -0.53 & -0.49  \\\hdashline
 			$\pi^{(2)}$ &  & 1 & -0.88 & -0.54 & 0.87 & 0.46  \\
 			$\pi^{(3)}$ &  &  & 1 & 0.41 & -0.96 & -0.32  \\
 			$\pi^{(\bar{3})}$ &  &  &  & 1 & -0.33 & -0.92  \\
 			$\pi^{(4)}$ &  &  &  &  & 1 & 0.27  \\
 			$\pi^{(\bar{4})}$ &  &  &  &  &  & 1  \\\hline\hline
 			
 			$\boldsymbol\Pi_{_W}$ & $X$ & $\pi^{(2)}$ & $\pi^{(3)}$ & $\pi^{(\bar{3})}$ & $\pi^{(4)}$ & $\pi^{(\bar{4})}$\\\hline
 			$F$ & 0.98 & -0.53 & 0.57 & 0.69 & -0.45 & -0.47 \\
 			$X$ & 1 & -0.59 & 0.66 & 0.67 & -0.53 & -0.45  \\\hdashline
 			$\pi^{(2)}$ &  & 1 & -0.89 & -0.53 & 0.93 & 0.46 \\
 			$\pi^{(3)}$ &  &  & 1 & 0.45 & -0.97 & -0.35 \\
 			$\pi^{(\bar{3})}$ &  &  &  & 1 & -0.39 & -0.95 \\
 			$\pi^{(4)}$ &  &  &  &  & 1 & 0.31  \\
 			$\pi^{(\bar{4})}$ &  &  &  &  &  & 1 \\\hline\hline
 			
 			$\boldsymbol\Pi_n$ & $X$ & $\pi^{(2)}$ & $\pi^{(3)}$ & $\pi^{(\bar{3})}$ & $\pi^{(4)}$ & $\pi^{(\bar{4})}$  \\\hline
 			$F$ & 0.98 & -0.48 & 0.56 & 0.57 & -0.44 & -0.50 \\
 			$X$ & 1 & -0.51 & 0.61 & 0.58 & -0.48 & -0.51  \\\hdashline
 			$\pi^{(2)}$ &  & 1 & -0.90 & -0.86 & 0.94 & 0.85  \\
 			$\pi^{(3)}$ &  &  & 1 & 0.90 & -0.96 & -0.87  \\
 			$\pi^{(\bar{3})}$ &  &  &  & 1 & -0.84 & -0.99 \\
 			$\pi^{(4)}$ &  &  &  &  & 1 & 0.82  \\
 			$\pi^{(\bar{4})}$ &  &  &  &  &  & 1  \\\hdashline
 			
 		\end{tabular}
 	\end{center}
 	\vspace{-10pt}
 	{\caption[Correlations between constant-maturity contract increments]{\small Correlations between 30-day constant-maturity risk premia increments based on daily, weekly and monthly monitoring over the full sample from January 1996 to December 2013.}\label{tab:swapcorrelations}}
 \end{table}
 As expected from the empirical study of \cite{DPS00} and many others since, the correlation between the daily changes in the S\&P500 forward and the variance swap in the top panel is around $-0.6$; the same holds for the correlation between the log contract and the variance swap. Both correlations decrease in magnitude, but only marginally, with the monitoring frequency, reaching the values $-0.48$ and $-0.51$ under monthly monitoring in the bottom panel, respectively. Thus, 
 variance swaps compensate the investor for downward shocks in the forward by a strongly positive realised variance.  The \gls{vrp} is negatively correlated with the third-moment and skewness risk premia and positively correlated with the fourth-moment and kurtosis premia at all monitoring frequencies. 
 Given its strong positive performance during crisis periods when large losses accrue to short variance swaps positions, the third-moment swap could even be attractive to variance swap issuers as a partial hedge.

Results for the skew and kurtosis risk premia are quite novel.\footnote{They are similar to the non-standardized third-moment and fourth-moment risk premia, respectively, so we confine our observations to skew and kurtosis.} There is a strong positive correlation between the forward  and the skew risk premium which increases with monitoring frequency: It is $0.74$ at the daily frequency but falls to $0.57$ at the monthly frequency. The correlations between the skew and kurtosis premia are strongly negative for all monitoring frequencies, ranging from $-0.92$ under daily monitoring to $-0.99$ under monthly monitoring for the standardised swaps. This indicates that skewness is clearly picking up the asymmetry in the tails of the distribution, rather than asymmetry around the centre. Under monthly monitoring the correlation of  $-0.86$ between the  variance and skew risk premiums is in line 
with \cite{KNS13}.\footnote{See p.13, Table 2, Panel B: The correlation between excess returns on the variance and cubic swap is $0.874$ where the positive sign comes from the fact that, in their setting, a writer of the cubic swap receives fixed an pays floating. The correlation between the variance and skewness swap is even stronger ($0.897$).} However, our more granular analysis allows for a more discerning conclusion, i.e. that standardised moment risk premia behave quite differently from their non-standardised counterparts when monitored at a higher frequency. The correlation between variance and the third-moment premiums remains almost as high at the daily frequency as it is at the monthly frequency (and similarly for the correlation between variance and the fourth moment). However, the correlation between the skew (kurtosis) premium and the \gls{vrp} decreases in magnitude from $-0.86$ ($0.85$) under monthly monitoring, to $-0.53$ ($0.46$) with weekly monitoring, and it remains at this level under daily monitoring.

\subsection{Determinants of Moment Risk Premia}\label{subsec:determinants}

Following the study  by \cite{CW09} on the determinants of variance risk premia, we now question whether significant common factors influencing our moment risk premia can be found among standard equity risk factors, namely: the excess return on the market ($ER$); the `small minus big' (\textit{size}) and the `high minus low' (\textit{growth}) factors introduced by \cite{FF93}; and the `up minus down' (\textit{momentum}) factor introduced by \cite{C97}. The size factor relates to the firm size and represents the historical excess returns of an investment in small firms over the investment in big firms. Historical excess returns of growth stocks over value stocks (as distinguished by the book-to-market ratio) are reflected in the growth factor. According to \cite{FF93} (p.4), these two factors also cover leverage and earnings-price-ratio effects. Finally, the momentum factor represents a momentum strategy and measures the excess returns of firms that performed well during the last time period over those who performed badly. 

Using monthly data on the S\&P500 \gls{vrp} from January 1996 through February 2003, \cite{CW09} find no significant effect for anything other than the market excess return as a driver of the \gls{vrp}. They also add a squared market factor as explanatory variable, as in the three-moment CAPM of \cite{KL76}, but find no evidence of an asymmetric response to market shocks. Our data construction methodology allows us to investigate the same phenomenon using higher frequency data. Given that \cite{E11} and  others document the importance of an asymmetric response in volatility  to market shocks at the daily frequency, it seems likely that daily or even weekly data would be sufficient to detect this effect. Using monthly data over the same period as \cite{CW09}, we also find no empirical evidence for an asymmetric response in the \gls{vrp}. 
However, using daily data over the same period the regression coefficient on the squared market factor is significantly different from zero at $0.1\%$. This finding leads us to question whether similar asymmetric responses are evident in third and fourth moment, and skewness and kurtosis risk premia, when measured at the daily frequency.

\begin{figure}[h!]
\begin{centering}
\includegraphics[width=1\linewidth,trim=3.65cm 0cm 2.6cm 0cm,clip=true]{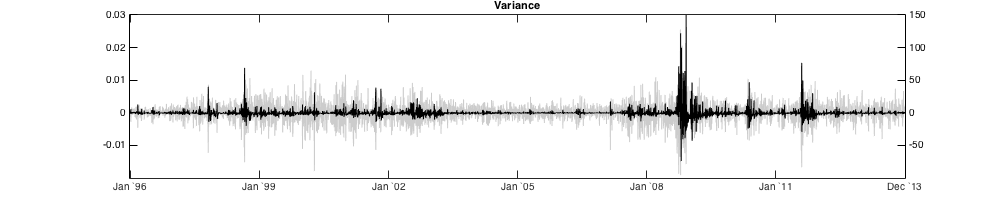}
\includegraphics[width=1\linewidth,trim=3.65cm 0cm 2.6cm 0cm,clip=true]{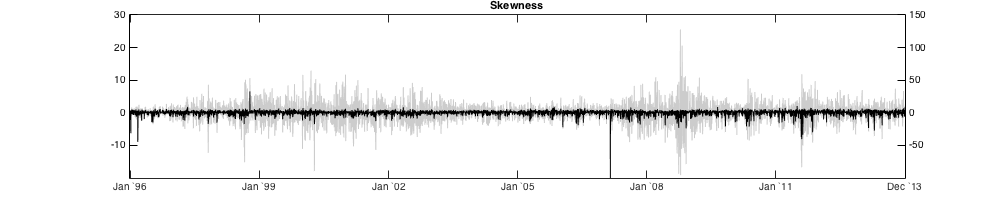}
\includegraphics[width=1\linewidth,trim=3.65cm 0cm 2.6cm 0cm,clip=true]{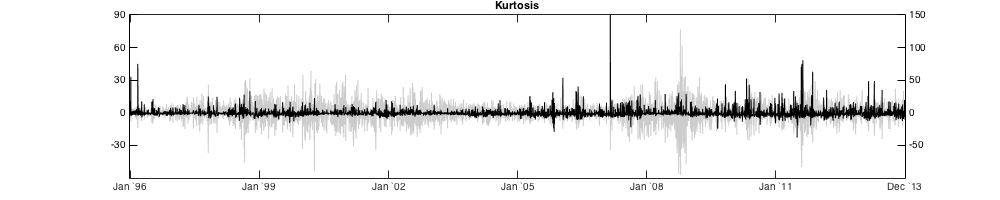}
\par\end{centering}
{ \caption[Moment futures increments]{\small Time series for the daily P\&L on 30-day variance, skewness and kurtosis swaps (black) and the S\&P500 futures (grey).}
\label{fig:increments}}
\end{figure}

 \noindent Figure \ref{fig:increments} presents time series on daily changes in the S\&P500 30-day constant-maturity synthetic futures price (in grey, measured on the right-hand scale) with a black line (measured on the left-hand scale) depicting daily changes in the 30-day, daily-monitored, DI \gls{vrp} (above), skew risk premium (middle) and kurtosis risk premium (below). The DI \gls{vrp} displays the well-known features common to the standard \gls{vrp}: it is typically small and negative but occasionally large and positive, in particular during the  a period of market turbulence. Notably, it has returned to small levels since the Eurozone crisis in August 2011 and during the credit boom period of mid-2003 to mid-2007. By contrast, the skew risk premium is typically small and positive, but occasionally takes large negative values. For instance, on 27 February 2007 it reached $-22.26$. On that day the S\&P500 index fell by 3.5\%, its biggest one-day fall since March 2003. The same day also marked a significant jump in the kurtosis risk premium, when it exceeded 160. Otherwise, like the \gls{vrp}, the kurtosis risk premium is usually small and negative. However, unlike the \gls{vrp}, the kurtosis risk premium has clearly increased in variability during the latter part of the sample.
 
 Following \cite{CW09} we now specify the regression model:\footnote{No significant autocorrelation is observed in the dependent and independent variables.  \cite{KNS13} do observe autocorrelation in daily data  -- on risk-neutral higher moments, as opposed to higher-moment risk premia -- but it is not sufficiently significant to be exploitable after transactions costs.}
\begin{equation}\label{eq:carhartregression}
\hat{\pi}=\alpha+\beta_{_{ER}}ER+\beta_{_{ER^2}}ER^2+\beta_{_{s}}\mbox{size}+\beta_{{g}}\mbox{growth}+\beta_{{m}}\mbox{momentum},
\end{equation}
where $\hat{\pi}$ denotes the daily change in the 30-day risk premium under consideration. Independent variables are measured using daily data on risk factors from the Fama-French website but also present results for a restricted model where $\beta_{{s}}=\beta_{{g}}=\beta_{{m}}=0$. We perform the analysis for the entire sample and separately for the financial crisis period between July 2008 and June 2009. We standardise all time series to make coefficients commensurate in size. As a result the intercept cannot be interpreted as an expected risk premium, but the beta coefficients can be interpreted as the number of standard deviations a risk premium is expected to change per standard deviation change in the corresponding factor.

\begin{table}[h!]\small
\vspace{10pt}
\begin{center}
\tabcolsep=0.12cm
\begin{tabular}{l|cc:cc:cc:cc:cc}
96-13 & \multicolumn{2}{c}{Variance} & \multicolumn{2}{c}{Third Moment} & \multicolumn{2}{c}{Skew} & \multicolumn{2}{c}{Fourth Moment} & \multicolumn{2}{c}{Kurtosis} \\ \hline
\multirow{2}{*}{$\alpha$} & -0.14 & -0.14 & 0.04 & 0.04 & 0.06 & 0.06 & -0.08 & -0.08 & -0.07 & -0.07 \\ 
& (-13.45) & (-13.57) & (3.73) & (3.74) & (5.76) & (5.71) & (-6.10) & (-6.10) & (-5.40) & (-5.17) \\ \hdashline
\multirow{2}{*}{$\beta_{_{ER}}$} & -0.61 & -0.63 & 0.65 & 0.65 & 0.70 & 0.74 & -0.49 & -0.50 & -0.48 & -0.52 \\ 
& (-62.73) & (-61.06) & (57.90) & (56.20) & (68.00) & (68.91) & (-39.97) & (-38.28) & (-37.71) & (-39.39) \\ \hdashline
\multirow{2}{*}{$\beta_{_{ER^2}}$} & 0.14 & 0.14 & -0.04 & -0.04 & -0.06 & -0.06 & 0.08 & 0.08 & 0.07 & 0.07 \\ 
& (42.33) & (42.38) & (-11.74) & (-11.67) & (-18.13) & (-17.84) & (19.21) & (19.07) & (17.01) & (16.16) \\ \hdashline
\multirow{2}{*}{$\beta_{{s}}$} & & 0.06 & & -0.07 & & -0.02 & & 0.05 & & -0.03 \\ 
& & (6.43) & & (-6.56) & & (-2.14) & & (4.10) & & (-2.21) \\ \hdashline
\multirow{2}{*}{$\beta_{{g}}$} & & -0.09 & & 0.12 & & 0.03 & & -0.11 & & -0.07 \\ 
& & (-8.89) & & (10.68) & & (2.43) & & (-8.51) & & (-4.86) \\ \hdashline
\multirow{2}{*}{$\beta_{{m}}$} & & 0.00 & & -0.05 & & 0.14 & & 0.05 & & -0.14 \\ 
& & (0.39) & & (-4.31) & & (12.79) & & (3.93) & & (-10.52) \\ \hline
$R^2$ & 0.567 & 0.581 & 0.439 & 0.469 & 0.528 & 0.544 & 0.309 & 0.331 & 0.280 & 0.298 \\ 
$F$ & \multicolumn{2}{c}{(50.1)} & \multicolumn{2}{c}{(83.8)} & \multicolumn{2}{c}{(56.0)} & \multicolumn{2}{c}{(50.9)} & \multicolumn{2}{c}{(40.0)} \\ 
\multicolumn{11}{c}{}\\
08-09 & \multicolumn{2}{c}{Variance} & \multicolumn{2}{c}{Third Moment} & \multicolumn{2}{c}{Skew} & \multicolumn{2}{c}{Fourth Moment} & \multicolumn{2}{c}{Kurtosis}\\ \hline
\multirow{2}{*}{$\alpha$} & -0.65 & -0.76 & 0.30 & 0.44 & 0.10 & 0.11 & -0.52 & -0.65 & -0.10 & -0.11 \\ 
& (-4.61) & (-5.66) & (2.01) & (3.22) & (2.88) & (2.91) & (-2.99) & (-3.95) & (-2.97) & (-3.09) \\ \hdashline
\multirow{2}{*}{$\beta_{_{ER}}$} & -1.04 & -1.24 & 1.40 & 1.69 & 0.37 & 0.36 & -1.23 & -1.53 & -0.23 & -0.23 \\ 
& (-18.64) & (-16.19) & (23.63) & (21.62) & (25.72) & (17.01) & (-17.60) & (-16.13) & (-16.94) & (-11.68) \\ \hdashline
\multirow{2}{*}{$\beta_{_{ER^2}}$} & 0.15 & 0.17 & -0.05 & -0.07 & -0.03 & -0.04 & 0.10 & 0.12 & 0.03 & 0.03 \\ 
& (11.84) & (13.76) & (-3.49) & (-5.35) & (-10.28) & (-10.35) & (6.16) & (7.89) & (9.66) & (9.87) \\ \hdashline
\multirow{2}{*}{$\beta_{{s}}$} & & 0.26 & & -0.22 & & -0.05 & & 0.22 & & 0.04 \\ 
& & (3.68) & & (-3.10) & & (-2.43) & & (2.46) & & (2.05) \\ \hdashline
\multirow{2}{*}{$\beta_{{g}}$} & & 0.30 & & -0.38 & & -0.00 & & 0.47 & & 0.02 \\ 
& & (3.16) & & (-3.94) & & (-0.04) & & (3.97) & & (0.73) \\ \hdashline
\multirow{2}{*}{$\beta_{{m}}$} & & -0.07 & & 0.13 & & -0.01 & & -0.07 & & 0.01 \\ 
& & (-0.72) & & (1.38) & & (-0.42) & & (-0.66) & & (0.22) \\ \hline
$R^2$ & 0.658 & 0.704 & 0.694 & 0.750 & 0.753 & 0.756 & 0.579 & 0.639 & 0.600 & 0.603 \\ 
$F$ & \multicolumn{2}{c}{(13.7)} & \multicolumn{2}{c}{(19.6)} & \multicolumn{2}{c}{(2.0)} & \multicolumn{2}{c}{(14.8)} & \multicolumn{2}{c}{(1.6)}\end{tabular}
{\caption[Time series for constant-maturity contracts]{\small Estimates and t-statistics (in brackets) as well as adjusted $R^2$ and F-test (in brackets) on joint significance for the restricted and unrestricted regression of the moment risk premia from January 1996 to December 2013 as well as for the crisis period from July 2008 to June 2009.}\label{tab:regression}}
\end{center}
\end{table}
Being based on more than 4500 observations, our analysis over the entire period provides some highly significant results.  The first blocks of both panels in Table \ref{tab:regression} report our results for the \gls{vrp}. The linear and quadratic excess return factors have highly significant loadings, the negative $\hat\beta_{_{ER}}$ being compensated by a positive $\hat{\beta}_{_{ER^2}}$. Thus the \gls{vrp} increases more when there is a negative market return than it decreases when there is a positive return of the same size. This asymmetric response is particularly pronounced during the financial crisis period (bottom panel). Over the whole 18-year period (top panel) the coefficients on the size and growth factors are small but significant, indicating that firm size has a positive impact and firm growth a negative impact on the \gls{vrp}, respectively. The addition of the Fama-French factors only marginally increases the adjusted $R^2$ from $0.567$ to $0.581$ but the $F$-statistic for addition of these factors is significant.

During the financial crisis the $R^2$ increases considerably relative to its value over the full sample as the \gls{vrp} becomes more sensitive to market shocks. The Fama-French factors, however, remain only marginally significant. Of these only the size factor has a significant coefficient of the same sign as for the full sample estimate. The change in sign of the coefficient on growth underlines the fact that July 2008 -- June 2009 represents a very particular market regime. The momentum factor appears to be irrelevant for both periods considered. 

The second column block of Table \ref{tab:regression} displays estimates for the third-moment risk premium. Here, the directional effects of the linear and quadratic factors are opposite to those observed in the variance premium regression: a market shock now has a greater impact on the third-moment premium when positive than when negative. The contribution of the size and momentum factors is relatively small but the growth factor has a significant positive effect which again changes sign during the financial crisis. Conclusions regarding the skewness premium (in the third column  block of the table) are similar, except that it is momentum rather than growth that has a positive effect on the skewness premium. It is remarkable that the explanatory power during the crisis period for the third moment is as high as $0.694$ ($0.750$ for the unrestricted model) and even higher ($0.753$) for the skewness risk premium. In fact, during the financial crisis the Fama-French and Carhart factors have almost no impact on the skewness premium: the F-statistic for joint significance is only $2.0$. 

The fourth and fifth column blocks of Table \ref{tab:regression} analyse the determinants of the fourth-moment and kurtosis risk premia. The much lower $R^2$ here indicates that these premia may be driven, to a large extent, by so far unknown risk factors. Otherwise the conclusions drawn are similar to -- yet weaker than -- those drawn about the variance premium. Apart from the excess  market return the only consistently significant effect is exhibited by the growth factor  for the fourth-moment risk premium and by the momentum factor  for the kurtosis risk premium, where the signs of the corresponding coefficients are opposite to those for the regression on the third-moment and skewness risk premia.

\section{Conclusions}\label{sec:conc}

We introduce a simple theory and a data-construction methodology which allows more precise analysis of the behaviour and determinants of variance and higher-moment risk premia than in most  other related research. Our empirical study brings into question some previous findings where data are either autocorrelated purely as an artefact of the construction methodology, or non-investable, or do not correspond to a risk premia at a constant maturity. Using 30-day constant-maturity data on DI swaps we construct risk premia for variance, third and fourth moment, skewness and kurtosis swaps and anayse their statistical properties. 
  
The \gls{vrp} is negatively correlated with the third-moment and skewness risk premia and positively correlated with the fourth-moment and kurtosis premia at all monitoring frequencies. There is a strong positive correlation between the forward  and the skew risk premium which increases with monitoring frequency. The correlations between the skew and kurtosis premia are strongly negative for all monitoring frequencies. This indicates that skewness is clearly picking up the asymmetry in the tails of the distribution, rather than asymmetry around the centre. 

 Our more granular analysis than in previous work shows that standardised moment risk premia behave quite differently from their non-standardised counterparts when monitored at a higher frequency. The correlation between variance and third-moment (and fourth-moment) premiums remains almost as high at the daily frequency as it is at the monthly frequency, but variance has a correlation with both skewness and kurtosis risk premia of about $-0.5$ and $0.5$, respectively, except under monthly monitoring, when the correlation of  $-0.86$ between the  variance and skew risk premiums is in line with \cite{KNS13}. 

We develop the study  by \cite{CW09} on the determinants of variance risk premia, using the same Fama-French factors, this time with a different data set, and results are presented over an 18-year period, with the banking crisis sub-sampled to check robustness. Our findings on the determinants of the VRP differ in that, using investable, 30-day constant-maturity data, we find a significant asymmetric response of the VRP to market shocks, as one would anticipate given the asymmetric volatility clustering that is well-documented in financial econometrics. We also find that size and growth are minor but significant drivers of the VRP, being positively and negatively related to the VRP, respectively.   

Determinants of the skewness and kurtosis (and third- and fourth- moment) risk premia have not previously been analysed. Our novel findings are that both have significant and asymmetric responses to market excess returns, and are also driven by momentum. Given that only size and growth are marginally significant determinants of the VRP, not momentum, this can explain the relatively low correlation that we observe between the VRP and the skewness risk premium, which contrast with the findings or \cite{KNS13}. Despite their high correlation, the much lower $R^2$ for the fourth-moment regressions indicates that the kurtosis premia may be driven, to a large extent, by so far unknown risk factors.

\newpage
\singlespacing
\bibliographystyle{plainnat}
\bibliography{bibliographyTRP}

\doublespacing
\begin{appendix}

\setlength\abovedisplayskip{6pt}
\setlength\belowdisplayskip{6pt}
\setlength\abovedisplayshortskip{0pt}
\setlength\belowdisplayshortskip{6pt}

\newpage

\section{Theoretical Appendix}\label{sec:app1}

\subsection{Proof of Theorem 1}\label{app:theorem2}

All characteristics of the form $\phi\left(\mathbf{\hat{x}}\right)=\text{tr}\left(\boldsymbol\Omega\mathbf{\hat{x}}\mathbf{\hat{x}}^\prime\right)$ are Discretisation-Invariant since
\begin{eqnarray*}
\mathbbm{E}\left[\lim_{\boldsymbol\Pi_N\rightarrow\boldsymbol\Pi}\sum_{\boldsymbol\Pi_{_N}}\text{tr}\left(\boldsymbol\Omega\mathbf{\hat{x}}_i\mathbf{\hat{x}}_i^\prime\right)\right]
&=&\mathbbm{E}\left[\text{tr}\left(\boldsymbol\Omega\lim_{\boldsymbol\Pi_N\rightarrow\boldsymbol\Pi}\sum_{\boldsymbol\Pi_{_N}}\left(\mathbf{x}_{t_i}-\mathbf{x}_{t_{i-1}}\right)\left(\mathbf{x}_{t_i}-\mathbf{x}_{t_{i-1}}\right)^\prime\right)\right]\\
&=&\mathbbm{E}\left[\text{tr}\left(\boldsymbol\Omega\lim_{\boldsymbol\Pi_N\rightarrow\boldsymbol\Pi}\sum_{\boldsymbol\Pi_{_N}}\left(\mathbf{x}_{t_i}\mathbf{x}_{t_i}^\prime-\mathbf{x}_{t_{i-1}}\mathbf{x}_{t_{i-1}}^\prime\right)\right)\right]\\
&=&\mathbbm{E}\left[\text{tr}\left(\boldsymbol\Omega\left(\mathbf{x}_{_T}\mathbf{x}_{_T}^\prime-\mathbf{x}_{_0}\mathbf{x}_{_0}^\prime\right)\right)\right]\\
&=&\mathbbm{E}\left[\text{tr}\left(\boldsymbol\Omega\left(\mathbf{x}_{_T}-\mathbf{x}_{_0}\right)\left(\mathbf{x}_{_T}-\mathbf{x}_{_0}\right)^\prime\right)\right],
\end{eqnarray*}
where the only requirement is that $\mathbf{x}$ follows a martingale (not necessarily an It\^o process). 
We can assume wlog that $\boldsymbol\Omega$ is a symmetric matrix because $\text{tr}\left(\boldsymbol\Omega\mathbf{\hat{x}}\mathbf{\hat{x}}^\prime\right)$ is a quadratic form.\qed

\subsection{Proof of Theorem 2}\label{app:theorem3}

With the value process of a $\phi$-swap being defined as
\begin{equation*}
\pi^\phi_t:=\mathbbm{E}_t\left[\sum_{\boldsymbol\Pi_{_N}}\phi\left(\mathbf{\hat{x}}_i\right)\right]-s^\phi_{_0},
\end{equation*}
the increments of the value process along the partition $\boldsymbol\Pi_{_N}$ are given by
\begin{eqnarray*}
\hat{\pi}^\phi_i&=&\pi^\phi_{t_i}-\pi^\phi_{t_{i-1}}=\mathbbm{E}_{t_i}\left[\sum_{\boldsymbol\Pi_{_N}}\phi\left(\mathbf{\hat{x}}_{\tilde{i}}\right)\right]-\mathbbm{E}_{t_{i-1}}\left[\sum_{\boldsymbol\Pi_{_N}}\phi\left(\mathbf{\hat{x}}_{\tilde{i}}\right)\right]\nonumber\\
&=&\sum_{\tilde{i}=1}^{i}\phi\left(\mathbf{\hat{x}}_{\tilde{i}}\right)+\mathbbm{E}_{t_i}\left[\sum_{\tilde{i}=i+1}^N\phi\left(\mathbf{\hat{x}}_{\tilde{i}}\right)\right]-\sum_{\tilde{i}=1}^{i-1}\phi\left(\mathbf{\hat{x}}_{\tilde{i}}\right)-\mathbbm{E}_{t_{i-1}}\left[\sum_{\tilde{i}=i}^N\phi\left(\mathbf{\hat{x}}_{\tilde{i}}\right)\right]\\
&=&\phi\left(\mathbf{\hat{x}}_i\right)+\mathbbm{E}_{t_i}\left[\phi\left(\mathbf{x}_{_T}-\mathbf{x}_{t_i}\right)\right]-\mathbbm{E}_{t_{i-1}}\left[\phi\left(\mathbf{x}_{_T}-\mathbf{x}_{t_{i-1}}\right)\right]\nonumber\\
&=&\phi\left(\mathbf{\hat{x}}_i\right)+\hat{s}^\phi_i
\end{eqnarray*}
where $\hat{s}^\phi_i=s^\phi_{t_i}-s^\phi_{t_{i-1}}$ and $s^\phi_t=\mathbbm{E}_t\left[\phi\left(\mathbf{x}_{_T}-\mathbf{x}_t\right)\right]$.
Combining this with the Theorem yields
\begin{eqnarray*}
\hat{s}^\phi_i&=&\mathbbm{E}_{t_i}\left[\text{tr}\left(\boldsymbol\Omega\left(\mathbf{x}_{_T}-\mathbf{x}_{t_i}\right)\left(\mathbf{x}_{_T}-\mathbf{x}_{t_i}\right)^\prime\right)\right]-\mathbbm{E}_{t_{i-1}}\left[\text{tr}\left(\boldsymbol\Omega\left(\mathbf{x}_{_T}-\mathbf{x}_{t_{i-1}}\right)\left(\mathbf{x}_{_T}-\mathbf{x}_{t_{i-1}}\right)^\prime\right)\right]\\
&=&\mathbbm{E}_{t_i}\left[\text{tr}\left(\boldsymbol\Omega\mathbf{x}_{_T}\mathbf{x}_{_T}^\prime\right)\right]-\text{tr}\left(\boldsymbol\Omega\mathbf{x}_{t_i}\mathbf{x}_{t_i}^\prime\right)-\mathbbm{E}_{t_{i-1}}\left[\text{tr}\left(\boldsymbol\Omega\mathbf{x}_{_T}\mathbf{x}_{_T}^\prime\right)\right]+\text{tr}\left(\boldsymbol\Omega\mathbf{x}_{t_{i-1}}\mathbf{x}_{t_{i-1}}^\prime\right)\\
&=&\text{tr}\left(\boldsymbol\Omega\boldsymbol{\hat{\Sigma}}_i\right)-\text{tr}\left(\boldsymbol\Omega\mathbf{x}_{t_i}\mathbf{x}_{t_i}^\prime\right)+\text{tr}\left(\boldsymbol\Omega\mathbf{x}_{t_{i-1}}\mathbf{x}_{t_{i-1}}^\prime\right)
\end{eqnarray*}
where $\boldsymbol{\hat{\Sigma}}_i=\boldsymbol{\Sigma}_{t_i}-\boldsymbol{\Sigma}_{t_{i-1}}$ with $\boldsymbol{\Sigma}_t=\mathbbm{E}_t\left[\mathbf{x}_{_T}\mathbf{x}_{_T}^\prime\right]$.
Thus
\begin{eqnarray*}
\hat{\pi}^\phi_i&=&\text{tr}\left(\boldsymbol\Omega\left(\mathbf{x}_{t_i}-\mathbf{x}_{t_{i-1}}\right)\left(\mathbf{x}_{t_i}-\mathbf{x}_{t_{i-1}}\right)^\prime\right)+\hat{s}^\phi_i\\
&=&\text{tr}\left(\boldsymbol\Omega\left[\boldsymbol{\hat{\Sigma}}_i-2\mathbf{x}_{t_{i-1}}\mathbf{\hat{x}}_i^\prime\right]\right).
\end{eqnarray*}
The fair-value swap rate becomes
\begin{eqnarray*}
s^\phi_{_0}&=&\mathbbm{E}\left[\phi\left(\mathbf{x}_{_T}-\mathbf{x}_{_0}\right)\right]\\
&=&\mathbbm{E}\left[\text{tr}\left(\boldsymbol\Omega\left(\mathbf{x}_{_T}-\mathbf{x}_{_0}\right)\left(\mathbf{x}_{_T}-\mathbf{x}_{_0}\right)^\prime\right)\right]\\
&=&\mathbbm{E}\left[\text{tr}\left(\boldsymbol\Omega\left[\mathbf{x}_{_T}\mathbf{x}_{_T}^\prime-\mathbf{x}_{_0}\mathbf{x}_{_0}^\prime\right]\right)\right]\\
&=&\text{tr}\left(\boldsymbol\Omega\left[\boldsymbol\Sigma_{_0}-\mathbf{x}_{_0}\mathbf{x}_{_0}^\prime\right]\right).
\end{eqnarray*}
This proves the Theorem.\qed

\end{appendix}
\end{document}